\documentclass[12pt,preprint]{aastex}

\shorttitle{WRs in IC10}
\shortauthors{Massey \& Holmes}

\begin{document}

\title{Wolf-Rayets in IC10: Probing the Nearest Starburst\altaffilmark{1}}

\author{Philip Massey, Shadrian Holmes\altaffilmark{2}}

\affil{Lowell Observatory, 1400 W. Mars Hill Road, Flagstaff, AZ 86001}
\email{Phil.Massey@lowell.edu, sholmes@astro.as.utexas.edu}

\altaffiltext{1}{Observations reported here were obtained at (1) the MMT Observatory, a joint facility of the University of Arizona and the
Smithsonian Institution, and (2) Kitt Peak National Observatory,
a division of the National Optical Astronomy Observatory, which is
operated by the Association of Universities for Research in Astronomy,
Inc., under cooperative agreement with the National Science Foundation.}
\altaffiltext{2}{Current address: Department of Astronomy, 
University of Texas at Austin, RLM 16.318 Austin, TX 78712-1083}

\begin{abstract}
IC10 is the nearest starburst galaxy, as revealed both by its H$\alpha$
surface brightness and the large number of Wolf-Rayet stars (WRs) per unit area.
The relative number of known WC- to WN-type WRs has been thought to be 
unusually high ($\sim 2$), unexpected for IC10's metallicity.  In this
{\it Letter} we report the first results of a new and deeper survey for
WRs in IC10.  We sucessfully detected all of the spectroscopically known WRs,
and based upon comparisons with a neighboring control field, estimate that
the total number of WRs in IC10 is about 100.  We present spectroscopic
confirmation of two of our WR candidates, both of which are of WN type.
Our photometric survey predicts that the actual WC/WN ratio is $\sim 0.3$.
This makes the WC/WN ratio of IC 10
consistent with that expected for its metallicity,
but greatly increases the already unusually high number of WRs, resulting
in a surface density that is about 20 times higher than in the LMC.  If the
majority of these candidates are spectroscopically confirmed, IC10 must
have an exceptional population of high mass stars. 

\end{abstract}


\section{Introduction}

Mayall (1935) first recognized IC10 as an extragalactic object, and, in
{\it The Realm of the Nebula,} Hubble (1936) proposed that it was likely
a member of the Local Group.  Its location just 3 degrees out of the 
Galactic plane has hampered investigations, but Hubble's description of it as 
``one of the most curious objects in the sky" has proven prophetic.
Today IC10 is understood to be an irregular galaxy undergoing an
intense burst of star formation likely triggered by infalling gas
from an extended cloud which is counter-rotating with respect to the
galaxy proper, as discussed by Wilcots \& Miller (1998), who conclude
that IC10 is a galaxy that is still forming.  

The starburst nature of IC10 was revealed primarily from the high number of
Wolf-Rayet stars (WRs) found by Massey, Conti \& Armandroff (1992) and 
Massey \& Armandroff (1995).  Hodge \& Lee (1990) had motivated these
studies by their discovery of 144 H~II regions, the brightest
of which were known to be comparable to the brightest seen in the SMC
(Hunter \& Gallagher 1985; Kennicutt 1988), 
a galaxy known to contain a substantial massive star population. 
Massey et al.\ (1992) used 
interference imaging to identify 22 WR candidates, of which 15 were
confirmed spectroscopically (Massey \& Armandroff 1995).  
This number was quite unexpectly high.  IC 10 is about half the size
of the SMC (van den Bergh 2000), which 
contains 11 WRs (Massey \& Duffy 2001); thus the overall surface density
of WRs in IC10 is at least 5 times greater than in the SMC. 
The galaxy-wide surface density of WRs in IC10 is in fact comparable to 
that of the most active OB associations in M~33 
(Massey \& Armandroff 1995).  The distribution of WRs across IC10 shows
that this high star-formation activity is not confined to a few regions
(which would simply be the result of statistical fluctuations or ``graininess"
in the star formation rate), but
rather is characteristic of a the galaxy as a whole.  
This is the classic definition of a starburst galaxy
(Hunter 1986; Searle \& Sargent 1972). 

\begin{figure}
\epsscale{1.0}
\plotone{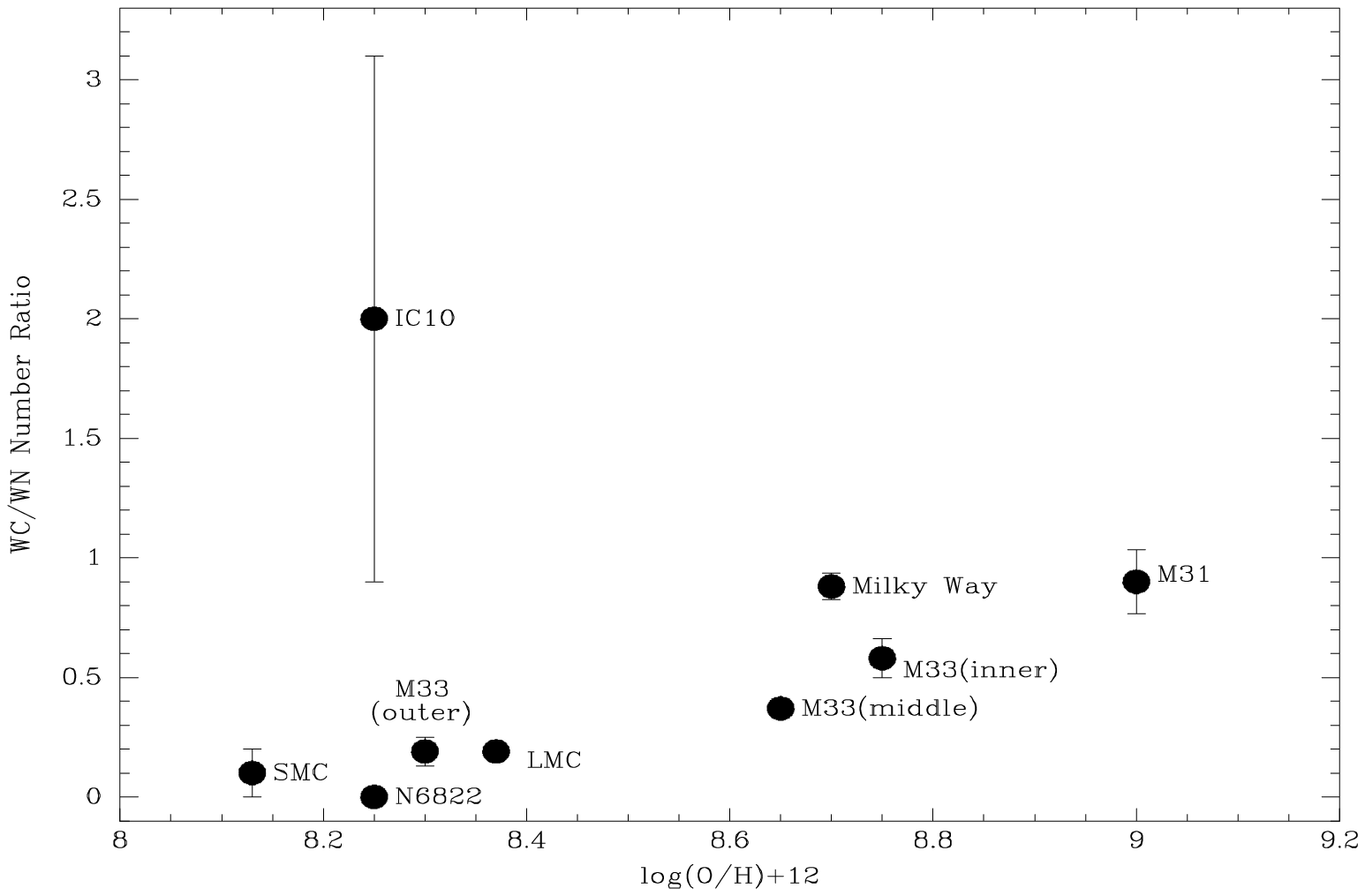}
\caption{\label{fig:wcwn} The
WC/WN number ratio is shown as a function of the oxygen
abundance. The data are from Massey \& Johnson (1998) and references
therein, updated for the
SMC from Massey \& Duffy (2001). The error bars are simply statistical; i.e.,
$\sigma_{\rm WC/WN}=\sqrt{({\rm WC/WN})^2 (1/{\rm WC} + 1/{\rm WN})}$,
except for NGC~6822 which contains no WCs.}
\end{figure}

However, one of the very peculiar results of these WR 
studies was the abnormally
large ratio of WC to WN stars given IC10's metallicity (log O/H+12=8.25,
Skillman, Kennicutt \& Hodge 1989; Garnett 1990).  Figure~\ref{fig:wcwn}
shows the relative number of WC and WN stars plotted for different galaxies
of the Local Group. The interpretation of the strong trend with metallicity
is straightforward: since the stellar winds of massive stars are driven
by radiation pressure in highly ionized metal lines, stars of a given
luminosity (and mass) will have a lower mass-loss rate in a lower metallicity
system, and hence will lose less mass during their lifetime.  In the
``Conti scenario" (Conti 1976, Maeder \& Conti 1994) a massive star
peels down like an onion due to mass-loss, revealing first the equilibrium
products of the CNO cycle at its surface (WN stars), and then the He-burning
products (WC stars).  Very massive stars will therefore evolve first to
the WN stage and subsequently to the WC stage, while a less massive star
might evolve only through WN.  At low metallicities the WCs should come
only from the very highest mass stars (Massey 2003).

The peculiar WC/WN ratio may be telling us something important about 
the star formation process in this, the nearest starburst galaxy.  If the IMF
was top-heavy (or inverted),
with an overabundance of the very highest mass stars, this
could explain the results.  However, Hunter (2001) finds a normal IMF
slope for the intermediate mass stars in IC10.  It would be very odd for the
IMF of the highest mass stars to be decoupled from that of the
intermediate-mass stars.  Alternatively, if the burst that produced the
Wolf-Rayet progenitors had been extremely coeval, then an abnormal
WC/WN population could certainly result (Schaerer \& Vacca 1998). However,
that would require a burst of duration less than 200,000 years over a scale
of a kpc.  Instead, a third, more prosaic possibility, is that 
the WC/WN ratio is strongly affected by incompleteness.  WN stars are
much harder to detect than WC stars as their strongest emission lines
are considerably weaker (see Massey \& Johnson 1998). 
Massey \& Armandroff (1995) argued against this possibility on the basis
that they had detected one WN star with lines as weak as commonly
associated with WNs.  However, since that time Royer et al.\ (2001, hereafter
RSMV) have
reported the discovery of 13 new 
WR candidates in IC10. (One of these,
their number 9, is actually
identical to star 6 in Massey et al.\ 1992.)  
Taken at face value, the additional stars would actually increase the WC/WN
ratio rather than decrease it.  
Spectroscopy of this sample by Crowther et al.\ (2002a) 
confirmed 9, leaving the WC/WN ratio little changed.
Nevertheless, the study certainly calls into
question the completeness of the original Massey et al.\ (1992) sample.

The purpose of this {\it Letter} is to report on the first results of a new,
much deeper search for Wolf-Rayet stars in IC10, 
along with spectroscopic confirmation
of two of these stars.  Complete details will be reported when
the spectroscopic followup is complete.  However, the number of good 
WR candidates found is striking, and probably resolves the issue of
the peculiar WC/WN ratio.

\section{The New Survey}
IC10 was imaged through three interference filters with the 4-m Mayall
telescope and Mosaic CCD camera (8k x 8k).  The data were obtained on
UT 19 and 20 Sept 2001.  The filter system is based upon that described by
Armandroff \& Massey (1985) and Massey \& Johnson (1998), but in large,
5.75~in by 5.75~in (146~mm $\times$ 146~mm)
format. The {\it WC} filter is centered at
C~III $\lambda 4650$, the strongest 
optical line in WC stars.  The {\it WN} filter is centered on 
He~II $\lambda 4686$, the strongest emission line in WN stars (although it
is also present in WCs; see Smith 1968).  
The continuum filter {\it CT} is centered at
$\lambda 4750$.  The central wavelengths were designed for use in the fast,
f/3 beam of the 4-m prime focus camera, and are roughly 50\AA\ in width.
The exposure time was 1.5 hrs in each filter, with the exposures broken
into three 1800 sec exposures with the telescope dithered by 150 arcsecs
NS and EW between exposures.
The scale is 0.27 arcsec pixel$^{-1}$, and the seeing
on the nine images ranged from 0.85 to 0.97 arcsec, with an average of 
0.92 arcsec. The transparency was excellent. (For comparison, 
the Massey et al.\ 1992 survey for WRs in IC10
used 1 hr exposures through each filter under 2.0-2.6 arcsec seeing with
drifting clouds interrupting the exposures.)  Exposures of spectrophotometric
standards through the {\it CT} filter were used to determine the continuum
magnitudes m$_{\lambda 4750}$, and are accurate to 0.1~mag.

The Mosaic camera consists of 8 separate chips, each covering 18.4 arcmin (EW)
by 9.2 arcmin (NS), each large compared to the optical extent of IC10
(half-light radius 2 arcmins).  IC10 had been centered on one of the chips,
and it and a neighboring chip were reduced in the identical manner.  The
latter was intended to serve as a control.
The data were processed through the IRAF Mosaic pipeline with
refinements from the \anchor{http://www.lowell.edu/~massey/lgsurvey/}
{Local Group Survey project.}  Instrumental
magnitudes were obtained using the point-spread-function fitting routine
``daophot", as implemented under IRAF.  All together, 114,000 stellar images
were photometered.  On average, 5300 stars were measured on each frame of
the control chip, and 7400 stars were measured on each frame of the galaxy
chip.

Candidate WRs were selected by comparing the magnitude differences {\it WC-CT}
and {\it WN-CT} to the uncertainty in the magnitudes based upon photon
statistics and read-noise, after a zero-point magnitude adjustment was
made based upon the full ensemble of stars.
Stars with magnitude differences more negative
than $-0.10$~mag and whose significance level was $>3\sigma$ were considered
valid candidates. The 9 frames were treated as 3 independent sets, grouped
by the three dithered telescope positions.  Each candidate was examined on
an image display by eye and checked for problems.  Altogether the search
revealed 238 unique
candidates on the galaxy field, and 135 unique
candidates for the control field; many of the candidates were found multiple
times.  We expect that none of the ``WR candidates" in the control field
are real, given their location $\sim 10$ arcmin from IC10.  Such spurious
detections are expected given the small magnitude differences we are looking
for, and the possible presence of absorption features in the {\it CT}
filter with non-WRs.  It is for this reason that we used a control field.

\section{Results}

The search found {\it all} 24 of the spectroscopically confirmed WRs
Massey \& Armandroff 1995; Crowther et al.\ 2002a) with
significance levels ranging from $5.6\sigma$ to $83\sigma$.  The weakest-lined
star had a magnitude difference of $-0.5$~mag, and the strongest-lined
star $-2.9$~mag between the WR filter ({\it WC} or {\it WN}) and the continuum
filter {\it CT}. P. Crowther (private communication) kindly conveyed the
specifics of which RSMV stars he had been able to confirm 
spectroscopically, and their spectral types in advance of publication.  
We therefore note that our survey
successfully distinguishes all of the known WCs from the known WNs if we adopt
a dividing line of {\it WC-WN}=$-0.1$.  Late-type WNs will have strong
N~III 4634,42 in the {\it WC} filter, while early-type WCs will have
very broad C~III $\lambda 4650$ spilling over into the {\it WN} filter,
so stars with a small absolute magnitude difference {\it WC-WN} are hard
to classify just based upon our filter photometry.

The Royer, Vreux, \& Manfroid (1998)
WR filter system uses five filters to help
classify WR candidates to excitation subtype (WN2, WN3 ... WN9; 
WC4, WC5 ... WC9).
Thus on the basis of their photometry alone, Royer et al.\ (2001) announced
the detection of very late-type WC (WC9) stars in IC10. This result was highly
surprising, as WC9 stars have previously been found only in much higher
metallicity environments, for reasons thought to be well understood from
stellar evolution: late-type WCs are thought to result from more enriched
surface material, and the star can only peel down far enough to
reveal these layers if the metallicity is high and mass-loss rates during
the O-type stages are therefore high (Smith \& Maeder 1991).  
Crowther et al.\ (2002b)
has recently called that into question, suggesting instead that the
late-type WCs are the result of
stronger stellar winds in the WR phase itself, but in either event WC9s
are not expected in low metallicity environments. This too was felt to be
part of the puzzle of star formation in IC10 (Royer et al.\ 2001).  
However, our survey detected
{\it none} of these WC9 candidates. 
An attempt to perform a quantitative spectroscopic analysis of these stars
by Crowther et al.\ (2002a) using GEMINI failed
to detect any emission.  We can probably conclude that these stars are not
real WRs. 

\begin{figure}
\epsscale{0.7}
\plotone{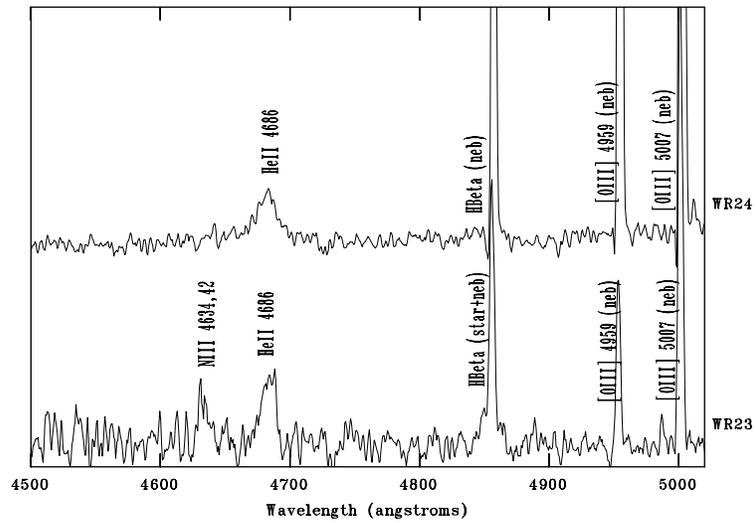}
\caption{\label{fig:spectra} The region around He~II $\lambda 4686$ is
shown for our two newly confirmed WR stars.  Both are of WN type. The
data have been slighly smoothed. The normalized spectrum of WR24 has been
scaled by a factor of 10; zero intensity is at the bottom of the figure.}
\end{figure}

Although the detection of all the known IC10 WRs, and the lack of 
detection of the spectroscopically rejected WR candidates gives us strong
confidence in our survey, there is no substitute for spectroscopic confirmation.
During a period of poor seeing at the MMT 6.5-m (14 Sept 2002)
we took time from our
main program and observed two of our new candidates using a single
slit setting.  We used the 800 line/mm grating on the Blue Channel
with a 2~arcsec slit to obtain 3.8\AA\ resolution spectra in the blue;
the exposure time was 2700 secs.
The spectra are shown in Fig.~\ref{fig:spectra}: both are Wolf-Rayet
stars of WN type. The coordinates are given in Table~1, along with the
equivalent width (ew) and full-width-at-half-maximum (fwhm)
of the He~II $\lambda 4686$ line.
The lines are sufficiently broad to rule out the
possibility that either star 
is an Of star, which might also show He~II and/or N~III emission.  
IC10-WR24 is by far the brightest WR found in IC10, and is
likely a blend of a WR star and another star.  A blend would also explain
the very weak emission (Table~1) combined with a normal line width. 

\begin{figure}
\epsscale{0.7}
\plotone{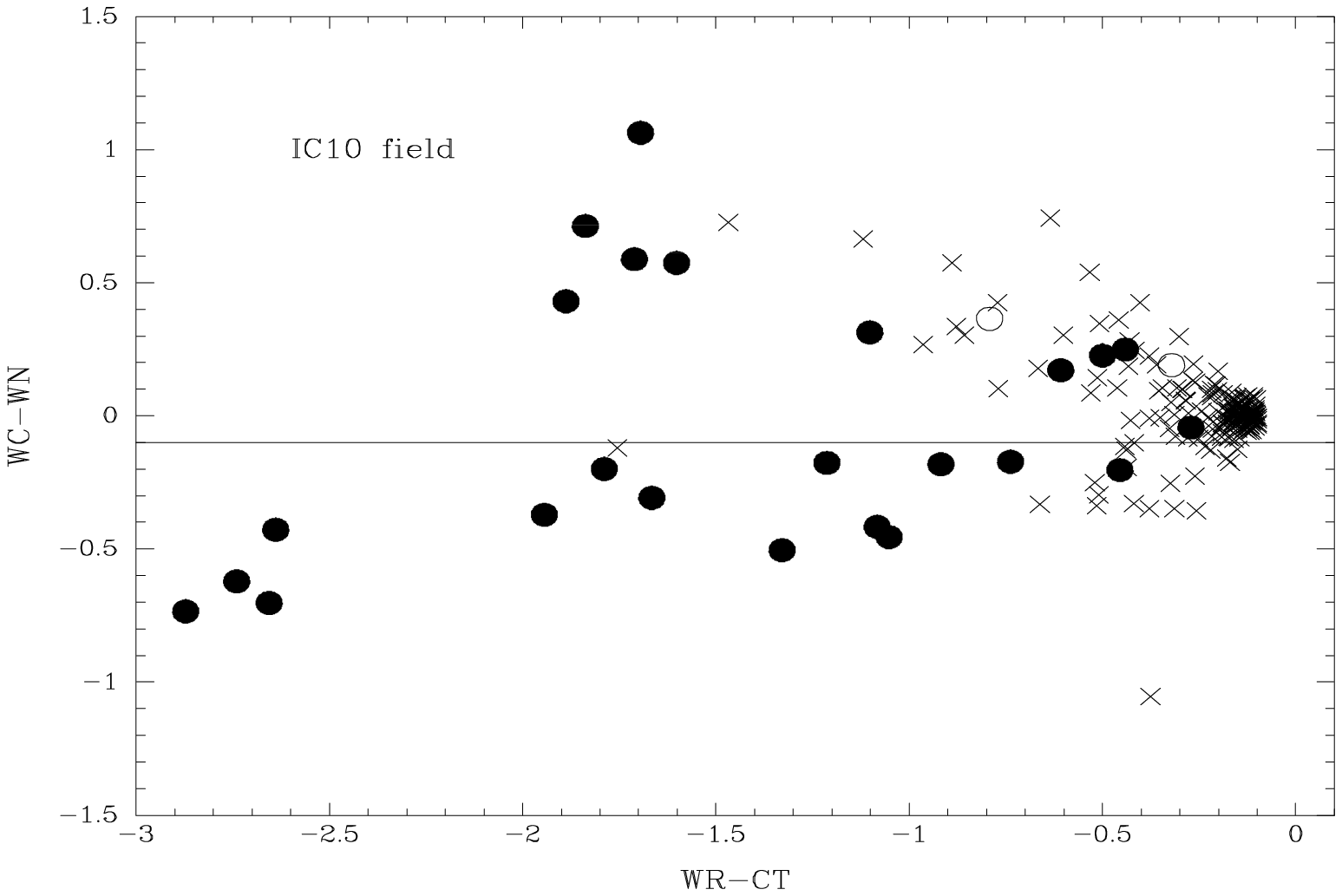}
\plotone{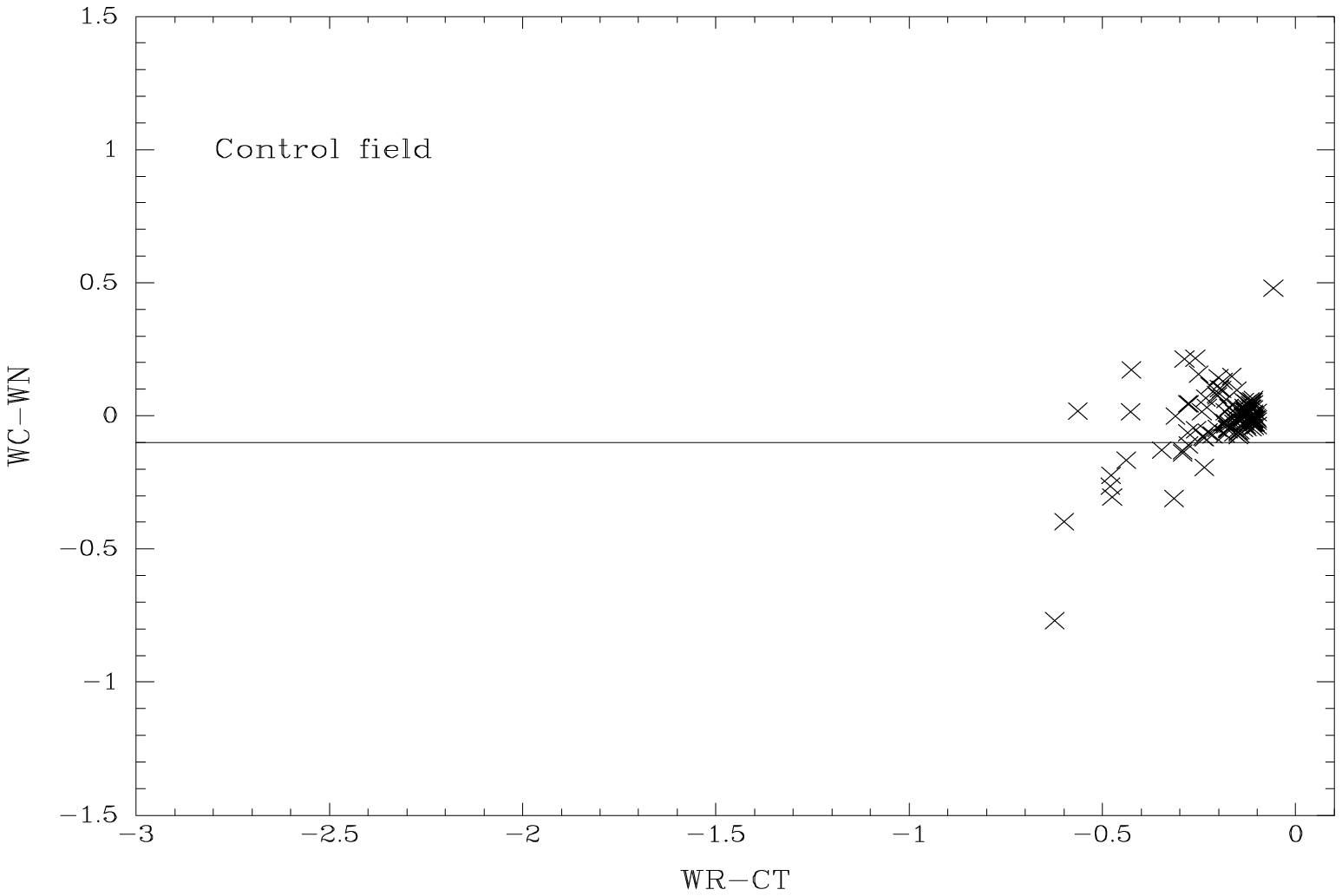}
\caption{\label{fig:compare} The results are our survey are shown both
for the IC10 field (upper) and control field (lower).  Crosses denote
new candidates; the filled circles are the stars that were previously
spectroscopically confirmed as WRs, and the two open circles denote
the stars newly confirmed here.  The magnitude difference {\it WC-CT}
or {\it WN-CT} (whichever was more negative) is shown as {\it WR-CT},
and is proportional to emission-line strength.  Stars above the line
(i.e., {\it WC-WN} $>-0.1$) are expected to be of WN type (if confirmed);
those below the line are expected to be of WC type. Note the lack
of strong-lined candidates in the control field.}
\end{figure}

In Figure~\ref{fig:compare} we show a comparison of the photometry of WR
candidates in the galaxy field with that from the control field.  We expect
that none of the 135 candidates in the latter are real.
Given that both the galaxy field (238 candidates) and control field
covered an equal area, we calculate that IC10 may
contain $\sim 100$ Wolf-Rayet stars in total.  Although this number seems
fantastically large, we note that 26 have now been confirmed spectroscopically:
15 by Massey \& Armandroff (1995), 9 by Crowther et al.\ (2002a),
and 2 here\footnote{This tally does not include the WN star reported by
Richer et al.\ (2001), as it may be coincident with RSMV~12; without better
identification it is impossible to tell. Including it would further lower
the WC/WN ratio and strengthen the argument presented here.}.

\begin{figure}
\epsscale{0.7}
\plotone{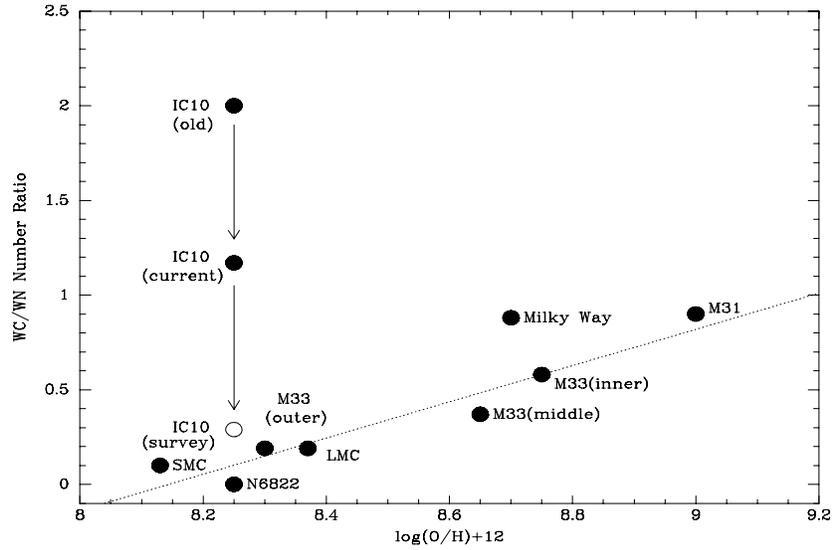}
\caption{\label{fig:wcwn2} The current WC/WN ratio of IC10 is still unusually
high based upon the newly confirmed WRs (Crowther et al.\ 2002a) and those
reported here.  However, if the number of new WRs found by our survey
is correct (when corrected for the number of false detections based upon
our control field) then the ratio becomes much more consistent with what
is expected on the basis of metallicity.  The dotted line is the least-squares
fit using the survey data for IC10, and ignoring the Milky Way, 
for which the data are probably incomplete, as discussed by Massey \& 
Johnson (1998).}
\end{figure}

What then can we conclude about the statistics of WCs and WNs in IC10?
First, for the spectroscopically confirmed WRs the WC/WN ratio is
now 1.2 rather than 2.0.  If we simply take all of the WR candidates in the 
IC10 field, and correct by the number of ``WC" and ``WN" detections in the
control field, we would expect to find a ratio of 0.3.  This ratio is only
slightly higher than that of the outer region of M~33, of similar metallicity
(Figure~\ref{fig:wcwn2}).  It is true that this result is somewhat dependent
upon our choice of the dividing line between ``WC" and ``WN" in our photometry.
While our choice is consistent with our knowledge of the IC10 WR spectral
types, spectroscopy of the remaining
candidates will be needed to confirm this result.

\section{Discussion}

If our statistical correction of the number of new candidates is correct,
then spectroscopy should be able to confirm an aditional
$\sim 2$ WCs and $\sim 66$ WNs in IC10.  Even
so, this may not represent the complete number, given the high reddening.
Thus the mystery of the high WC/WN ratio in IC10 may be solved.

However, spectroscopic confirmation of such large additional
number of WRs in IC10
would certainly make this galaxy even more unique in terms of its massive
star population.  Two pieces of evidence suggest that this may well be
the case.
First, of the spectroscopically confirmed WNs, two are
of WN7-8 type.
At low metallicities the only similar late-type WNs are found in the 30~Dor
region of the LMC, where very high mass stars abound.
Studies of coeval regions containing these stars in the Milky Way suggest
that they come from only the highest mass stars (Massey, DeGioia-Eastwood,
\& Waterhouse 2001), and we would expect the progenitors to be even more
massive in a low metallicity environment (Massey 2003).  
This is consistent then with IC10 having a normal IMF
but an exceptionally large
population of massive stars.  Secondly, the integrated H$\alpha$ emission
suggests that IC10 has one of the two highest rates of star formation per
unit area known of a representative sample of non-interacting irregular
galaxies (Hunter 1997 and private communication).

The ``active" area of IC10 is approximately 8 arcmin $\times$ 8 arcmin
in angular extent; at a distance of 660 kpc (Sakai, Madore, \& Freedman 1999), 
this correspond to an area of 2.2 kpc$^{2}$.  Thus if our estimate is correct,
IC10 would contain roughly 45 Wolf-Rayet stars kpc$^{-2}$.  For comparison,
the LMC contains $\sim 2$ WRs kpc$^{-2}$ (Massey \& Johnson 1998).  A typical
Galactic OB association might contain several WRs, and be 100~pc in
diameter, i.e., with a surface density of a couple of hundred WRs kpc$^{-2}$---
only several times larger than what we see {\it globally} in IC10.  Thus,
if confirmed, the high number of WR stars would suggest
that IC10 has a population of massive stars similar 
to that of an OB association but on a kpc scale.

\acknowledgments
We are thankful to Deidre Hunter for useful discussions, and to Paul Crowther
for communicating the results of his spectroscopy.
This work has been supported by the NSF through grant AST-0093060.

\clearpage

\clearpage

\clearpage

\clearpage

\clearpage
\begin{deluxetable}{cccccccl}
\tabletypesize{\scriptsize}
\tablecolumns{8}
\tablenum{1}
\tablecaption{Spectrophotometry of Newly Confirmed IC10 WRs}
\tablehead{
\colhead{} &
\colhead{} & 
\colhead{} &
\colhead{} &
\colhead{} &
\multicolumn{2}{c}{He~II $\lambda 4686$} & 
\colhead {}  \\ \cline{6-7}
\colhead{Star\tablenotemark{a}} &
\colhead{$\alpha_{\rm 2000}$} &
\colhead{$\delta_{\rm 2000}$} &
\colhead{$m_{\lambda 4750}$} &
\colhead{Type} &
\colhead{EW (\AA)} &
\colhead{FWHM (\AA)} &
\colhead{Comment} 
}
\startdata
23 & 00:20:32.79 & 59:17:16.4 & 22.3 & WN7-8 & -40 & 13 & hydrogen, strong NIII\\
24 & 00:20:27.73 & 59:17:37.2 & 18.8 & WN   & -4 & 21 &  \\
\enddata
\tablenotetext{a}{The numbering is a continuation of those of Massey et al. 1992
and Massey \& Johnson 1998. Revised designations including the ``RMSV" stars
will be included once spectroscopy of the new candidates is completed.}
\end{deluxetable}

\end{document}